\documentclass[11pt]{article}
\usepackage{svg}
\usepackage[margin=1in]{geometry}
\usepackage{amsmath,amssymb,amsthm}
\usepackage{graphicx}
\usepackage{booktabs}
\usepackage{siunitx}
\usepackage{microtype}
\usepackage{hyperref}
\usepackage{caption}
\usepackage{subcaption}
\usepackage{float}

\hypersetup{
  colorlinks=true,
  linkcolor=blue,
  citecolor=blue,
  urlcolor=blue
}

\newcommand{\Z}{\mathbb{Z}}
\newcommand{\MI}{\mathrm{I}}
\newcommand{\CV}{\mathrm{CV}}

\newcommand{\CPSK}{\mathrm{CPS}_K}

\title{Hardware Validation of DAGI via a Modular ``Ridge'' Signature and High-Order Synergistic Information}
\author{
  \textsc{Petr Sramek} \\
  \textit{Whytics, Cambridge, MA, USA} \\
  \textit{DAGI Research Program -- Empirical Validation Track} \\
  \texttt{p.sramek@whytics.com}
}
\date{Preprint v1.0: March 2026 \\ \small (Hardware Execution: December 2025)}

\begin{document}
\maketitle

\begin{abstract}
We report a hardware validation of the DAGI (Directed Acyclic Graph Information) framework on IBM Quantum hardware using a small, controlled experiment whose ideal output distribution is constrained to a low-dimensional modular manifold (a ``ridge''). For two $n$-bit registers $(u,v)$ with $n=4$ (modulus $16$), each key instance $k$ induces an ideal relation $v \equiv k\,u \ (\mathrm{mod}\ 16)$, producing a visually distinct ridge in the joint $(u,v)$ distribution. Executed on \texttt{ibm\_torino} in a single SamplerV2 job (8 keys, 1024 shots/key, $N=8192$ total shots), the ridge persists under hardware noise with ridge-hit probability $p_{\mathrm{hit}}=0.1830$ (uniform baseline $1/16$), corresponding to a ridge contrast of $2.93\times$ (95\% bootstrap CI $[2.80,3.06]$). Key recovery exceeds chance: per-shot accuracy $0.1689$ (chance $0.125$, 95\% Wilson CI $[0.1610,0.1772]$), and per-group dictionary recovery $0.375$ (chance $0.125$). 
To test the central DAGI hypothesis---that recoverable key information is predominantly \emph{high-order/synergistic} rather than visible in low-order marginals---we compute a Möbius-based information decomposition of $\MI(K;D_S)$ over detector-bit subsets $S$ via a M\"obius inversion pipeline and evaluate targeted positive synergy $\CPSK$ at order $k_{\max}=3$. We observe $\CPSK(k=3)=0.08788$ with significance under label-shuffle permutation tests (accuracy $p=0.001996$, $\CPSK$ $p=0.004975$). Uniformity diagnostics show near-uniform single-bit marginals while correlation concentrates in specific low-order pairs, and a bootstrap reliability sweep confirms order-3 targeted synergy remains statistically reliable at the full 1024-shot target budget. These results support the claim that DAGI detects and quantifies nontrivial, hardware-resilient, higher-order information structure associated with a known global algebraic constraint.
\end{abstract}

\section{Introduction}

Modern quantum devices exhibit complex, correlated noise, and the observables accessible to hardware are often indirect and noisy views of latent algorithmic structure. A recurring practical problem is therefore: \emph{given only classical measurement bitstrings from hardware, how can we detect, quantify, and localize meaningful structure that survives noise?}

The DAGI framework addresses this by combining (i) information-theoretic observables (typically mutual information between a label/latent variable and subsets of measured bits), (ii) Möbius inversion on the subset lattice to decompose information across interaction order, and (iii) reliability controls (e.g., bootstrap-based frontiers) to avoid over-interpreting high-order terms when sampling noise dominates.

The DAGI (Directed Acyclic Graph Information) framework used here is developed in a set of core
manuscripts that (i) introduce the DAG-based ontology for informational branching and record formation,
and (ii) formalize the multiscale M\"obius-inversion machinery used to decompose $I(K;D_S)$ across subset
order and to quantify irreducible (synergistic) structure.
The foundational DAG-based interpretation and the M\"obius-enhanced multiscale formalism underlying the
metrics reported in this paper are described in
Refs.~\cite{SramekDAGI_Branching2025,SramekDAGI_Mobius2025}.
The present work should be read as a hardware validation of those definitions and diagnostics on noisy
quantum measurement data.

In this paper we validate DAGI on a deliberately simple but diagnostically sharp hardware target: a small Fourier/oracle-style circuit family whose ideal output distribution concentrates near a modular manifold (a ``ridge'') in the joint outcome space $(u,v)$. This provides a known \emph{global algebraic signature} against which we can test two claims:

\begin{enumerate}
  \item \textbf{Ridge survival under hardware noise:} the low-dimensional ridge remains visible on quantum hardware for small $n$.
  \item \textbf{High-order key information:} information about the key $k$ is not reducible to single-bit marginals, but appears as higher-order (synergistic) structure that DAGI can quantify and validate statistically.
\end{enumerate}

\section{Experimental Overview}

\subsection{Ridge relation and observable variables}

We measure two $n$-bit registers $(a,b)$ with $n=4$, interpret them (little-endian) as integers
\begin{equation}
u = \mathrm{int}(a) \in \Z_{16}, \qquad v = \mathrm{int}(b) \in \Z_{16},
\end{equation}
and evaluate a key-dependent ridge residual
\begin{equation}
r_k(u,v) \;=\; (v - k\,u) \bmod 2^n.
\end{equation}
The idealized signature for a given key $k$ is concentration of probability mass near the ridge
\begin{equation}
v \equiv k\,u \pmod{16}
\quad \Leftrightarrow \quad
r_k(u,v)=0.
\end{equation}

We consider a small key set
\begin{equation}
K = \{1,3,5,7,2,4,8,12\} \subset \Z_{16},
\end{equation}
chosen to include both odd and even keys (to probe aliasing/coset behavior under modulus $2^n$ while keeping the total hardware cost low).

\subsection{Hardware run and dataset}

All key instances were executed as separate circuits inside a single IBM Runtime SamplerV2 job on the \texttt{ibm\_torino} backend:

\begin{itemize}
  \item Backend: \texttt{ibm\_torino}
  \item Job ID: \texttt{d541drnp3tbc73amunv0}
  \item Registers: $n=4$ (8 measured bits total)
  \item Keys: $|K|=8$
  \item Shots per key: 1024
  \item Total shots: $N = 8192$
  \item Transpilation: optimization level 1; AQFT disabled
\end{itemize}

Transpiled circuit depths ranged from 334 to 476, with two-qubit gate counts from 188 to 253 (mean depth 411.1; mean 2Q count 224.5).

\section{Metrics and Statistical Tests}

\subsection{Ridge hit and ridge contrast}

Define the ridge-hit indicator for a given key $k$:
\begin{equation}
L_k(u,v) = \mathbf{1}[r_k(u,v)=0].
\end{equation}
Let $p_{\mathrm{hit}}$ denote the empirical ridge-hit probability under the true key label. Under a null model where $(u,v)$ is uniform over $\Z_{16}\times\Z_{16}$, the ridge-hit probability is exactly $2^{-n}=1/16$.

We define ridge contrast as
\begin{equation}
\mathrm{Contrast} \;=\; \frac{p_{\mathrm{hit}}}{2^{-n}}.
\end{equation}

\subsection{Key recovery metrics}

We report two complementary key-recovery metrics:
\begin{enumerate}
  \item \textbf{Per-shot key accuracy:} a classifier predicts $\hat{k}(u,v)$ from each individual outcome, and accuracy is $\Pr[\hat{k}=k]$.
  \item \textbf{Per-group dictionary recovery:} for each true key $k$, aggregate its 1024 shots and select the key that maximizes ridge-hits (or minimizes ridge distance) across that group. This corresponds to a simple ``dictionary attack'' style recovery and is reported as group accuracy over the 8 key groups.
\end{enumerate}

Chance accuracy for both metrics is $1/|K| = 0.125$.

\subsection{DAGI / Möbius information decomposition}

Let $D=(D_1,\dots,D_{2n})$ denote the measured bits (here $2n=8$). Let $K$ denote the key label (discrete, 8 classes). For any subset of measured bits $S\subseteq \{1,\dots,2n\}$, define
\begin{equation}
g(S) \;=\; \MI(K; D_S),
\end{equation}
the mutual information between the key label and the subset of measured bits.

We compute a Möbius inversion on the subset lattice to obtain a decomposition
\begin{equation}
g(S) \;=\; \sum_{T \subseteq S} f(T),
\end{equation}
where $f(T)$ can be interpreted as the \emph{irreducible} (non-overlapping) contribution attributable to subset $T$. Positive $f(T)$ at $|T|\ge 3$ indicates \emph{synergistic} information about $K$ that requires jointly observing the bits in $T$.

We summarize by order $k$ via the \emph{positive mass}:
\begin{equation}
M_k^{+} \;=\; \sum_{|T|=k} \max\{f(T),0\}.
\end{equation}
We report the targeted order-3 positive synergy score
\begin{equation}
\CPSK(k=3) \;\equiv\; M_3^{+},
\end{equation}
computed with $k_{\max}=3$.

We further partition subsets by whether they lie \emph{within} a register or are \emph{cross-register} (touching at least one bit from each of $a$ and $b$), and report within/cross decompositions of positive mass.

\subsection{Reliability frontier}

To ensure higher-order estimates are not artifacts of finite sampling, we evaluate a bootstrap/subsampling reliability sweep over shots per key. For each shots/key budget, we estimate a coefficient of variation $\CV_k$ for $M_k^{+}$ (or its bootstrap estimate) and define the \emph{max reliable order} $k^\star$ as the largest $k$ satisfying
\begin{equation}
\CV_k \le 1.
\end{equation}

\subsection{Permutation tests}

We test whether the observed key accuracy and $\CPSK$ could arise under a null model by shuffling key labels across shots (label-shuffle permutation). The empirical $p$-value is computed as the fraction of permutations with statistic at least as extreme as observed.

\section{Results}

\subsection{Visual ridge survival on hardware}

\begin{figure}[H]
  \centering
  \includegraphics[width=\linewidth]{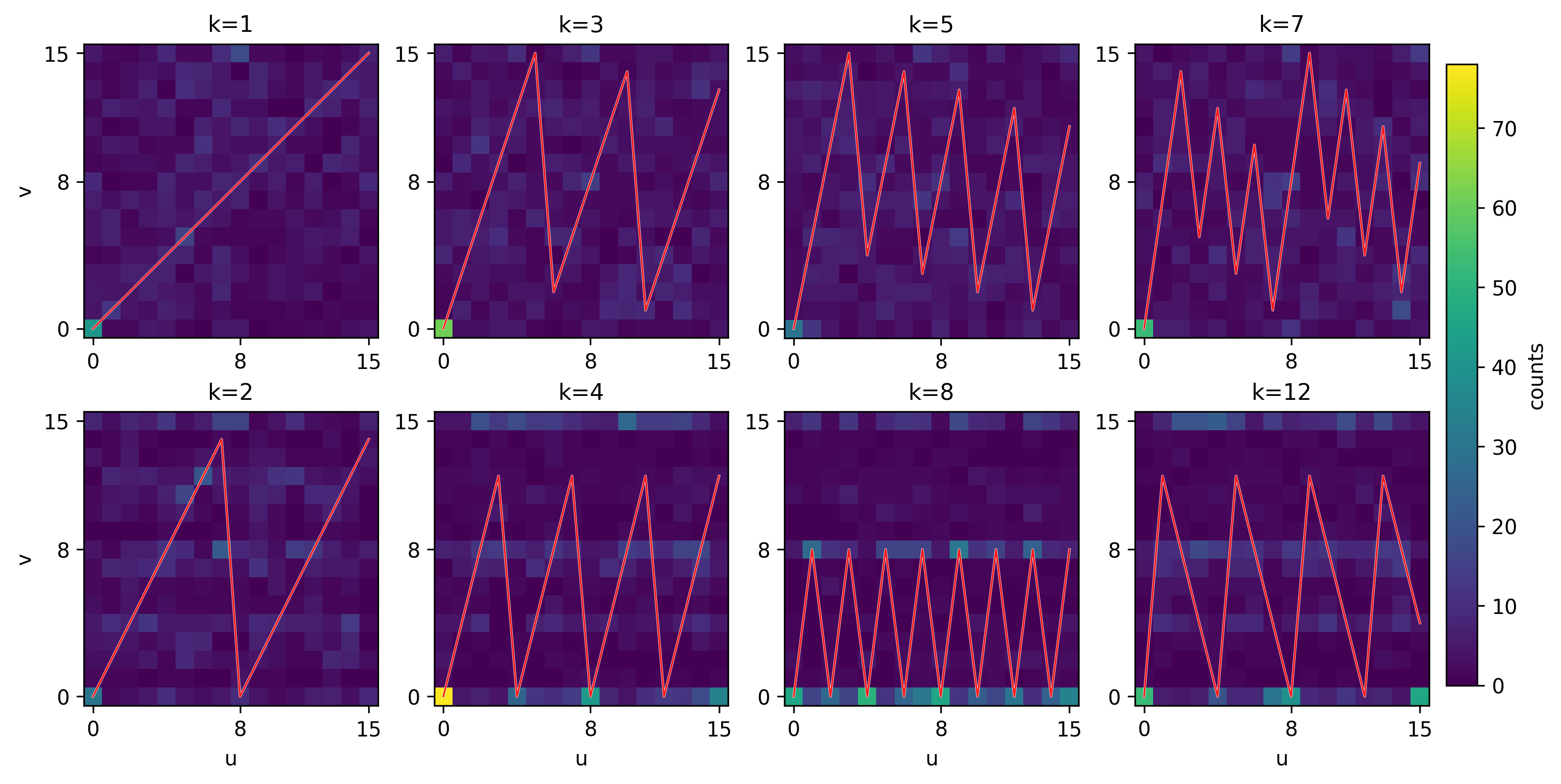}
  \caption{\textbf{Hardware ridge heatmaps.}
  16$\times$16 histograms of $(u,v)$ on \texttt{ibm\_torino} for $n=4$, 8 keys, 1024 shots/key ($N=8192$). Red overlay shows the predicted ridge $v\equiv k\cdot u\ (\mathrm{mod}\ 16)$ used in evaluation. Takeaway: the low-dimensional ridge structure is visibly present on hardware across keys.}
  \label{fig:ridge_heatmaps}
\end{figure}

Figure~\ref{fig:ridge_heatmaps} shows $16\times 16$ histograms of $(u,v)$ on \texttt{ibm\_torino}, with the predicted ridge $v\equiv k u \pmod{16}$ overlaid in red. The ridge is visibly present across keys, supporting survival of the low-dimensional algebraic signature under hardware noise.

\subsection{Headline quantitative metrics}

Table~\ref{tab:headline} reports the main metrics, including uncertainty for ridge-hit probability and key accuracy (Wilson intervals) and ridge contrast (bootstrap CI). The ridge-hit probability is $p_{\mathrm{hit}}=0.1830$, compared to the uniform baseline $1/16=0.0625$, corresponding to a $2.93\times$ ridge contrast. Per-shot key accuracy is $0.1689$, exceeding chance $0.125$.

\begin{table}[H]
  \centering
  \caption{\textbf{Headline outcomes on \texttt{ibm\_torino}.}
  $n=4$, $|K|=8$, 1024 shots/key, $N=8192$ total shots.}
  \label{tab:headline}
  \begin{tabular}{lrr}
    \toprule
    Metric & Value & 95\% CI \\
    \midrule
    Ridge hit $p$ & 0.1830 & [0.1748, 0.1915] \\
    Key accuracy (shot) & 0.1689 & [0.1610, 0.1772] \\
    Ridge contrast & 2.93 & [2.80, 3.06] \\
    $\CPSK$ (k=3) & 0.08788 & --- \\
    \bottomrule
  \end{tabular}
\end{table}

We also observe per-group dictionary recovery (one prediction per key group using all 1024 shots) of $0.375$ (3/8 correct), exceeding chance $0.125$ despite substantial aliasing between specific keys.

\subsection{Permutation significance of accuracy and high-order synergy}

\begin{figure}[H]
  \centering
  \includegraphics[width=\linewidth]{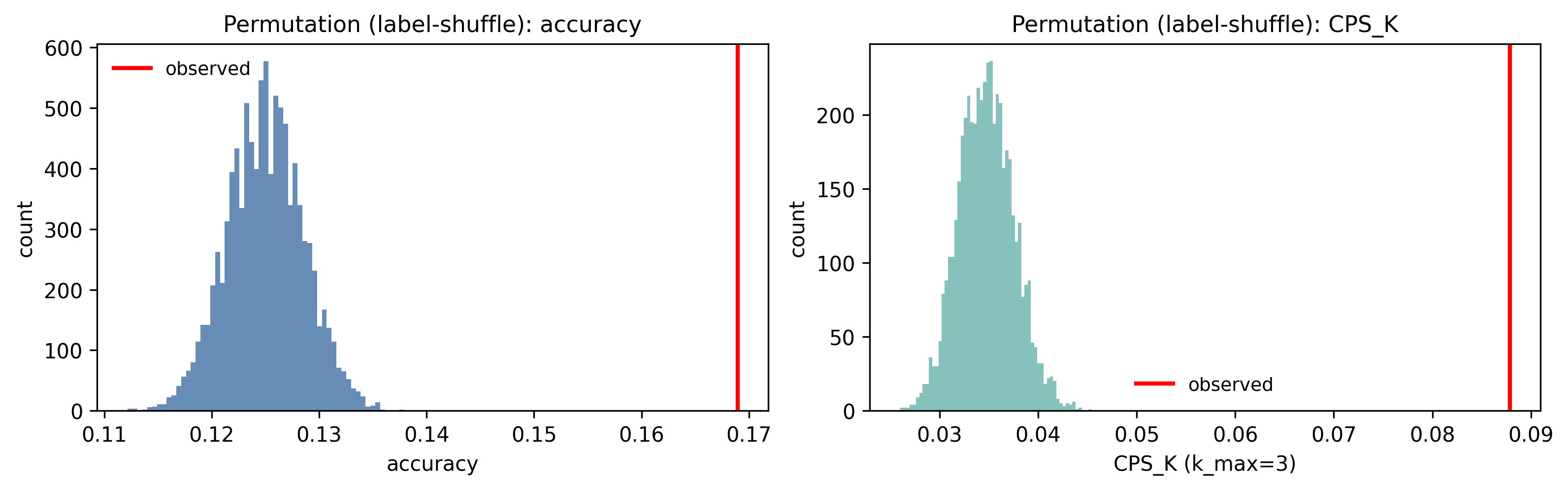}
  \caption{\textbf{Permutation tests (label shuffle).}
  Label-shuffle permutation distributions for per-shot key accuracy (500 perms) and $\CPSK$ (200 perms) computed on the same dataset ($N=8192$). Red line marks the observed value. Takeaway: both accuracy and $\CPSK$ are far into the tail of the null, supporting nontrivial key information.}
  \label{fig:permutation}
\end{figure}

Figure~\ref{fig:permutation} summarizes label-shuffle permutation tests for per-shot accuracy and $\CPSK$. In both cases, the observed statistics lie deep in the tail of the null distribution, supporting nontrivial key information and nontrivial higher-order structure. Using the permutation runs reported in the evaluation suite, we obtain:
\begin{equation}
p_{\mathrm{perm}}(\text{accuracy}) = 0.001996, \qquad
p_{\mathrm{perm}}(\CPSK) = 0.004975.
\end{equation}

\subsection{Uniformity sanity checks: marginals near-uniform, structure in low-order pairs}

A critical threat to interpretation is that apparent predictability could be driven by trivial single-bit bias. Uniformity diagnostics show that single-bit marginals are close to $0.5$ in the pooled dataset (pooled max marginal deviation $\approx 0.0724$), while correlation structure concentrates in specific low-order pairs, especially within the $b$-register. Figure~\ref{fig:uniformity} summarizes these checks and highlights the top absolute covariance pairs.

\begin{figure}[H]
  \centering
  \includegraphics[width=\linewidth]{fig3_uniformity_and_top_cov.png}
  \caption{\textbf{Uniformity and correlation structure.}
  Uniformity sanity checks on pooled marginals and top $|\mathrm{cov}|$ pairs (tagged within-A/within-B/cross). Dashed line marks $0.5$ for marginals. Takeaway: single-bit marginals are near-uniform while correlation structure concentrates in specific low-order pairs.}
  \label{fig:uniformity}
\end{figure}

\subsection{Targeted synergy is predominantly cross-register at order 3}

The DAGI decomposition yields positive mass by order (orders 1--3) and a within-vs-cross breakdown relative to the $(a,b)$ register partition. For order 3, the positive mass is overwhelmingly cross-register:
\begin{equation}
M_3^{+} = 0.08788,\quad
M_{3,\mathrm{cross}}^{+} = 0.08535,\quad
M_{3,\mathrm{within}}^{+} = 0.00253,
\end{equation}
i.e., $\approx 97\%$ of order-3 positive mass is cross-register. This is consistent with the ridge relation $v \equiv k u$ being fundamentally a \emph{cross-register} constraint.

We additionally evaluated an ``odd-key-only'' slice as a control, obtaining $\CPSK(k=3)\approx 0.0523$ (still positive), indicating that the high-order signal is not exclusively driven by even-key aliasing effects.

\subsection{Reliability frontier vs shots per key}

\begin{figure}[H]
  \centering
  \includegraphics[width=\linewidth]{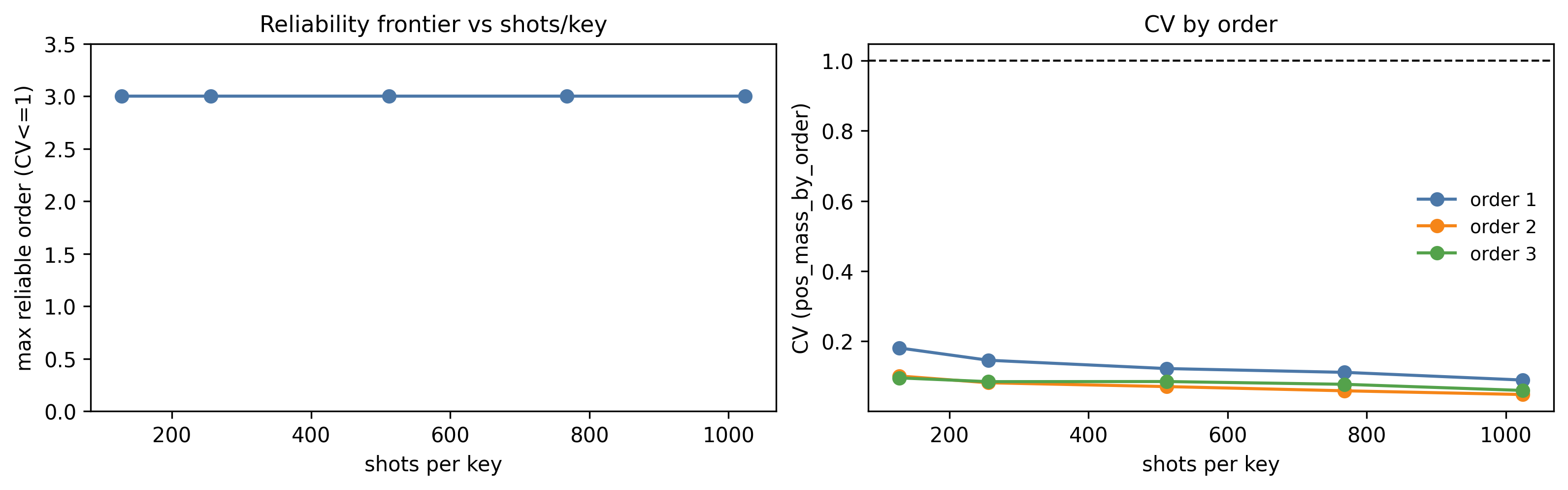}
  \caption{\textbf{Reliability frontier sweep.}
  Reliability sweep via bootstrap over shots/key $\in \{128,256,512,768,1024\}$, reporting $\CV$ of positive mass by order and max reliable order ($\CV\le 1$). Takeaway: order-3 targeted synergy remains statistically stable at the primary experimental budget (1024 shots/key).}
  \label{fig:reliability}
\end{figure}

Figure~\ref{fig:reliability} presents a bootstrap reliability sweep over shots/key $\in\{128,256,512,768,1024\}$. Using the criterion $\CV\le 1$, the max reliable order remains $k^\star=3$ throughout the sweep, and the order-3 targeted synergy remains statistically stable at the primary experimental budget (1024 shots/key).

\subsection{Ablation: higher-order representations improve predictability and/or calibration}

To connect the DAGI claim to predictive utility, we evaluated simple ablations that restrict the representation of each shot:
(i) marginals-only features, (ii) pairwise features, and (iii) full bitstring models, using a stratified 50/50 train/test split (4096/4096). Pairwise features provide the strongest test accuracy and best calibration (lowest ECE), indicating that structure beyond marginals is needed, and that naive ``use everything'' models can be miscalibrated in this regime.

\begin{figure}[H]
  \centering
  \includegraphics[width=\linewidth]{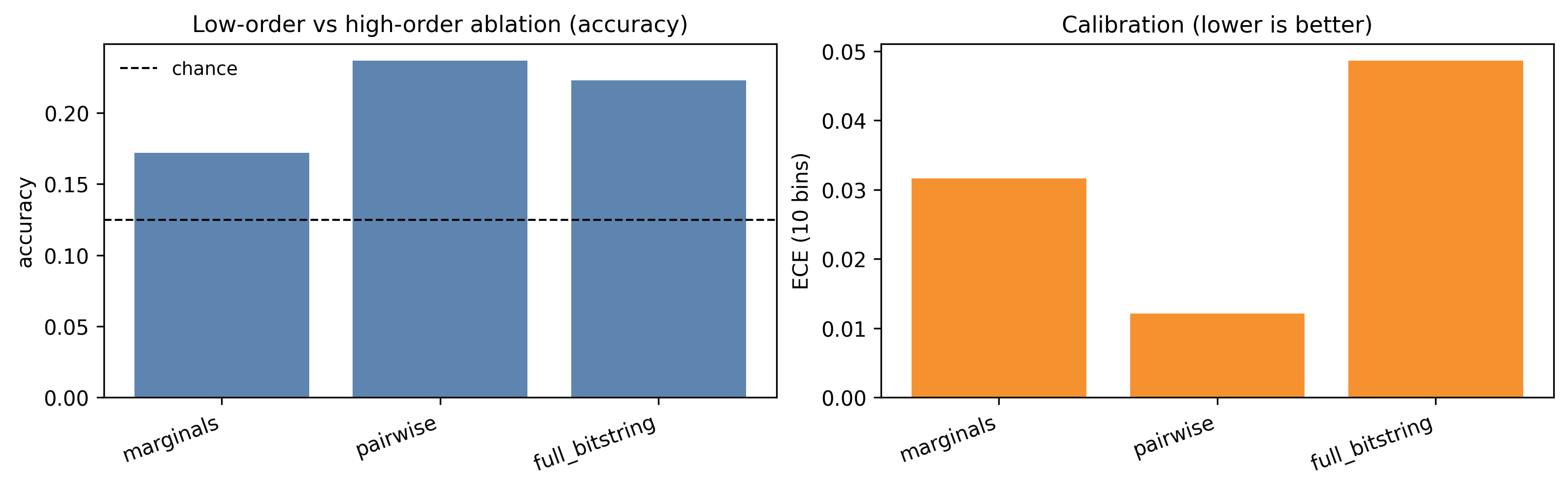}
  \caption{\textbf{Ablation and calibration.}
  Low-order vs high-order ablation using (i) marginals-only, (ii) pairwise features, and (iii) full-bitstring models, evaluated on a stratified 50/50 split. Takeaway: higher-order representations improve predictability and/or calibration beyond low-order summaries.}
  \label{fig:ablation}
\end{figure}

\begin{table}[H]
  \centering
  \caption{\textbf{Ablation summary (test set).} Chance accuracy is $1/8=0.125$. ECE is expected calibration error.}
  \label{tab:ablation}
  \begin{tabular}{lcc}
    \toprule
    Model & Test accuracy & Test ECE \\
    \midrule
    Marginals-only & 0.1721 & 0.0316 \\
    Pairwise features & 0.2368 & 0.0121 \\
    Full bitstring & 0.2231 & 0.0486 \\
    \bottomrule
  \end{tabular}
\end{table}

\section{Discussion}

\subsection{What this validates about DAGI}

This experiment was designed so that the ground-truth structure is not ambiguous: the correct key class determines a modular ridge constraint linking $u$ and $v$. The hardware results support two core claims:

\paragraph{(1) A low-dimensional manifold signature survives hardware noise.}
The ridge contrast of $2.93\times$ over uniform and the visible ridge in Fig.~\ref{fig:ridge_heatmaps} show that the global algebraic signature survives a realistic transpiled circuit on \texttt{ibm\_torino} at modest depth.

\paragraph{(2) Key information is not purely low-order; higher-order structure is statistically real.}
Uniformity diagnostics show marginals are near-uniform (Fig.~\ref{fig:uniformity}), yet label-shuffle tests confirm nontrivial predictability and nontrivial $\CPSK$ (Fig.~\ref{fig:permutation}). The order-3 positive mass is predominantly cross-register, matching the semantics of the ridge constraint. The ablation further supports the point that representations beyond marginals are required for strong predictability and calibration (Fig.~\ref{fig:ablation}, Table~\ref{tab:ablation}).

\subsection{Why we did not run the full 6-bit instance (and what that means)}
A full 6-bit (12 measured bits) instance would increase circuit depth and two-qubit gate counts substantially, and would also increase the sample complexity required for reliable estimation of high-order terms. The present validation therefore uses a reduced instance ($n=4$) chosen to preserve the ridge geometry and key-dependence while remaining hardware-feasible within a strict quantum-time budget.

This choice trades away a direct ``full-scale'' replication of the 6-bit demonstration reported in our companion Page-Time Synergy Peak validation \cite{Sramek2026PageTime}, but it does \emph{not} undercut the central validation logic: the experiment still instantiates a known global algebraic constraint that (i) produces a ridge signature and (ii) induces cross-register, higher-order information about the key. The measured survival of these effects on hardware constitutes a meaningful validation of DAGI in the intended sense: detecting and quantifying nontrivial structure that is invisible (or weak) in low-order marginals.

\subsection{Limitations and scope}
\begin{itemize}
  \item \textbf{Small $n$ and few keys:} this is a controlled validation, not a scaling claim. The appropriate next step is a tiered sweep in $n$ and circuit depth, with careful control of sample complexity and reliability frontiers.
  \item \textbf{Key aliasing under modulus $2^n$:} even keys can induce degeneracies in modular multiplication constraints; we partially address this with an odd-only slice that remains synergistic, but a more systematic study is warranted.
  \item \textbf{Model dependence of per-shot key recovery:} accuracy depends on the chosen per-shot scoring/classifier. The DAGI result is more robust in that it targets information structure rather than a single classifier family.
\end{itemize}

\section{Conclusion}

We presented a hardware validation of DAGI on IBM \texttt{ibm\_torino} using a modular ridge experiment with $n=4$ and 8 key instances. The ridge survives hardware noise with clear contrast over uniform, key recovery exceeds chance, and a Möbius/DAGI decomposition of $\MI(K;D_S)$ reveals statistically significant order-3 synergistic structure ($\CPSK(k=3)=0.08788$) that is predominantly cross-register and remains reliable under bootstrap sweeps. Together, these findings support the claim that DAGI can detect and validate nontrivial, hardware-resilient, high-order information structure in quantum measurement data.

\section*{Reproducibility and Artifact Manifest}

Hardware configuration and primary artifacts (provided in the supplementary data bundle accompanying this preprint on Zenodo):
\begin{itemize}
  \item Backend: \texttt{ibm\_torino}
  \item Hardware job ID: \texttt{d541drnp3tbc73amunv0}
  \item Dataset: Supplementary Zenodo bundle (\texttt{PlanIBM2112\_ecc\_ridge\_n4\_8keys\_1024.npz})
  \item Evaluation summary: Supplementary Zenodo bundle (\texttt{suite.json})
\end{itemize}

\section*{Acknowledgments}
This work used IBM Quantum services. The views expressed are those of the authors and do not reflect the official policy or position of IBM or the IBM Quantum team.

\end{document}